\documentclass[twocolumn,showpacs,preprintnumbers,amsmath,amssymb,prl]{revtex4}

\usepackage{graphicx}
\usepackage{bm}

\begin{document}

\title{The "True" Widom Line for a Square-Well System}

\author{V. V. Brazhkin}
\affiliation{Institute for High Pressure Physics, Russian Academy
of Sciences, Troitsk 142190, Moscow Region, Russia}

\author{Yu. D. Fomin}
\affiliation{Institute for High Pressure Physics, Russian Academy
of Sciences, Troitsk 142190, Moscow Region, Russia}

\author{V. N. Ryzhov}
\affiliation{Institute for High Pressure Physics, Russian Academy
of Sciences, Troitsk 142190, Moscow Region, Russia}

\author{E. E. Tareyeva}
\affiliation{Institute for High Pressure Physics, Russian Academy
of Sciences, Troitsk 142190, Moscow Region, Russia}

\author{E. N. Tsiok}
\affiliation{Institute for High Pressure Physics, Russian Academy
of Sciences, Troitsk 142190, Moscow Region, Russia}

\date{\today}

\begin{abstract}
In the present paper we propose the van der Waals-like model,
which allows a purely analytical study of fluid properties
including the equation of state, phase behavior and supercritical
fluctuations. We take a square-well system as an example and
calculate its liquid - gas transition line and supercritical
fluctuations. Employing this model allows us to calculate not only
the thermodynamic response functions (isothermal compressibility
$\beta_T$, isobaric heat capacity $C_P$, density fluctuations
$\zeta_T$, and thermal expansion coefficient $\alpha_T$), but also
the correlation length in the fluid $\xi$. It is shown that the
bunch of extrema widens rapidly upon departure from the critical
point. It seems that the Widom line defined in this way cannot be
considered as a real boundary that divides the supercritical
region into the gaslike and liquidlike regions. As it has been
shown recently, the new dynamic line on the phase diagram in the
supercritical region, namely the Frenkel line, can be used for
this purpose.
\end{abstract}

 \maketitle


In recent years, a growing attention has been given to the
investigation of properties of supercritical liquids. This
interest is mainly due to the fact, that supercritical fluids are
widely used in industrial processes. Their behavior away from the
critical point is therefore an important practical question
because it might affect their applicability in the considered
technological process \cite{appl_supp}. Theoretical aspects of the
physics of supercritical fluid are of particular interest as well.

The liquid-gas phase equilibrium curve in the $P-T$ plane ends at
the critical point. At pressures and temperatures above the
critical ones ($P > P_c$ and $T > T_c$), the properties of a
substance in the isotherms and isobars vary continuously, and it
is commonly said that the substance is in its supercritical fluid
state, when there is no difference between liquid and gas. From
the physical point of view, the $P,T$ region near the critical
point, where anomalous behavior of the majority of characteristics
is observed (the so-called critical behavior) is of prime interest
\cite{stanley_book}. The correlation length of thermodynamic
fluctuations diverges at the critical point \cite{stanley_book}.
One can also observe a critical behavior of the thermodynamic
response functions, which are defined as second derivatives of the
corresponding thermodynamic potentials; such as the
compressibility coefficient $\beta_T$, thermal expansion
coefficient $\alpha_P$, and heat capacity $C_P$. These quantities
pass through their maxima during pressure or temperature
variations and diverge as the critical point is approached. Near
the critical point, the positions of the maxima of these values in
the $T,P$-plane are close to each other. The same is true for the
density fluctuations, the speed of sound, thermal conductivity,
etc. Therefore, in the supercritical region, there is the whole
set of the lines of extrema of various thermodynamic parameters.
The lines of the maxima for different response functions
asymptotically approach one another as the critical point is
approached, because all response functions can be expressed in
terms of the correlation length. This asymptotic line is sometimes
called the "Widom line" and is often regarded as an extension of
the coexistence line into the "one-phase region" \cite{pnas2005}.
The Widom lines for the gas-liquid and liquid-liquid phase
transitions have been investigated extensively
\cite{pnas2005,poole,fr_st,mcm_st,bryk1,bryk2,jpcb,br_jcp,br_ufn,gallo_jcp,riem,may}.

Because of the lack of a theoretical method for constructing the
Widom line, based on the extremum of the correlation length, the
locus of extrema of the constant-pressure specific heat $C_P$ was
often used as an estimation for the Widom line. In Refs.
\cite{jpcb,br_ufn}, using the computer simulations, the locus of
extrema (ridges) for the heat capacity, thermal expansion
coefficient, compressibility, and density fluctuations for model
particle systems with Lennard-Jones (LJ) potential in the
supercritical region have been obtained. It was found that the
ridges for different thermodynamic values virtually merge into a
single Widom line at $T < 1.1T_c$ and $P < 1.5P_c$ and become
almost completely smeared at $T < 2.5T_c$ and $P < 10P_c$, where
$T_c$ and $P_c$ are the critical temperature and pressure. The
analytical expressions for the extrema of the heat capacity,
thermal expansion coefficient, compressibility, density
fluctuation, and sound velocity in the supercritical region were
obtained in Refs. \cite{br_jcp,br_ufn} in the framework of the van
der Waals (vdW) model. It was found that the ridges for different
thermodynamic values virtually merge into a single Widom line only
at $T <1.07T_c, P < 1.25P_c$ and become smeared at $T <2T_c, P <
5P_c$. However, in both of the studies, the estimation of the
Widom line is unsatisfactory because it is a priori unclear how
far from the critical point the lines of extrema of response
functions follow the exact Widom line, determined by the maximum
of the correlation length.

Recently it was proposed to construct the Widom line by using the
approach based on Riemannian geometry \cite{riem}. Previously it
was supposed that there is a relation between the Riemannian
thermodynamic scalar curvature $R$ of the thermodynamic metric and
the volume of the correlation length $\xi$, i.e., $|R|\propto
\xi^3$ \cite{riem2}. Consequently, the locus of the maximum of
$|R|$ describes the locus of the Widom line. In Ref. \cite{riem}
this approach was applied to a van der Waals (vdW) fluid, while in
\cite{may} it was used for constructing the Widom line for the LJ
fluid.

It is of great interest to develop a simple vdW-like model which
can represent a liquid-gas transition and the Widom line, and can
be solved analytically. In this case it would be possible to
analyze the relation between the"true" Widom line, determined from
the correlation function, and the lines of extrema of response
functions.


We define the model using the approximation for the direct
correlation function \cite{book} of the hard-core system,
suggested by Lovett \cite{lovett} (see also \cite{RT1,RT2,TMF}):
\begin{equation}
c(r)=\left\{
\begin{array}{ll}
c_{HS}(r) , & r\leq d \\
-\frac{\phi(r)}{k_BT}, & r>d
\end{array}%
\right.,   \label{1}
\end{equation}
where $c_{HS}(r)$ is the hard spheres direct correlation function
and $\phi(r)$ is the attractive part of the potential. This should
be a good approximation when $-\frac{\phi(r)}{k_BT}$ is small. The
approximation, though rough, is similar in spirit to the mean
spherical model approximation which has been found to be a good
approximation in many cases \cite{book}. Such an approximation
formulated directly in terms of $c(r)$ is particularly convenient
for the formulation of the Widom line, as it will be shown below.
This approximation is especcially convenient for direct
calculation of the correlation length in the fluid $\xi$.

Although the method we use is a general tool for liquids, we
consider the so-called square well (SW) system as more specific.
One can easily generalize the results to other systems. The square
well system is a system of particles interacting via the following
potential:
\begin{equation}
\Phi(r)=\left\{
\begin{array}{lll}
\infty , & r\leq d \\
-\varepsilon , & d <r\leq \sigma  \\
0, & r>\sigma%
\end{array}%
\right.. \label{2}
\end{equation}

Although this system is not very realistic, it can serve as a
generic example of a simple liquid. Below we consider the SW
system with $\sigma =1.35d$. In this case, Eq.~(\ref{1}) has the
form:
\begin{equation}
c(r)=\left\{
\begin{array}{lll}
c_{HS}(r) , & r\leq d \\
\frac{\varepsilon}{k_BT}, & d <r\leq \sigma  \\
0, & r>\sigma%
\end{array}%
\right.  \label{3}
\end{equation}

We use the Percus-Yevick approximation \cite{book} for $c_{HS}$:
\begin{equation}
c_{HS}(r)=- \lambda_1 - \pi \rho \lambda_2 r - \frac{\pi}{12}
  \rho \lambda_1 r^3, \label{py}
\end{equation}
where
\begin{equation}
  \lambda_1=\frac{(1+2 \eta)^2}{(1-\eta)^4} \nonumber
\end{equation}
and
\begin{equation}
  \lambda_2=\frac{-d^2(1+1/2 \eta)^2}{(1-\eta)^4}. \nonumber
\end{equation}
$\eta$ is the packing fraction $\eta=\frac{\pi}{6}\rho d^3$.

The direct correlation function can be used to obtain the
isothermal compressibility:
\begin{equation}
  \frac{1}{k_BT} \left( \frac{\partial P}{\partial \rho}
  \right)_T = 1- \rho \int d\textbf{r} c(r). \label{comp1}
\end{equation}

Taking into account, that the system is isotropic, and using
Eq.~(\ref{py}), one obtains:
\begin{equation}
  \frac{1}{k_BT} \left( \frac{\partial P}{\partial \rho}
  \right)_T = \lambda_1 -\frac{8 \varepsilon}{k_BT} \eta \Delta,
  \label{comp2}
\end{equation}
where $\Delta= (\sigma^3 - d^3)/d^3$.

The integration of Eq.~(\ref{comp2}) gives the equation of state
(EoS) for the SW system:
\begin{equation}
  \tilde{P}=\tilde{T} \cdot \frac{\eta + \eta^2 + \eta^3}{(1- \eta)^3} - 4
  \Delta \eta^2, \label{eos}
\end{equation}
where $\tilde{P}=P \cdot \frac{\pi d^3}{6 \varepsilon}$ and
$\tilde{T}=k_BT/ \varepsilon$. Further in the paper we will use
only these scaled units, omitting the tilde mark.

\begin{figure}
\includegraphics[width=8cm]{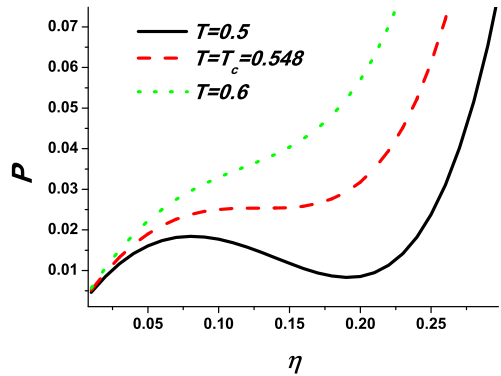}%

\caption{\label{fig:fig1} (Color online) Three isotherms for SW
system: below the critical temperature $T_c$, at $T=T_c$ and above
$T_c$.}
\end{figure}

Fig.~\ref{fig:fig1} shows three isotherms for the SW system: below
the critical temperature $T_c$, at $T=T_c$ and above $T_c$. The
critical point can be determined from the following conditions:
\begin{equation}
\frac{\partial P}{\partial \eta}= T \cdot \frac{(1+2 \eta)^2}{(1-
\eta)^4} - 8 \Delta \eta
\end{equation}
and
\begin{equation}
\frac{\partial ^2 P}{\partial \eta ^2}= T \cdot \frac{8+20 \eta +8
\eta ^2}{(1- \eta)^5} - 8 \Delta.
\end{equation}

Rewriting these equations in the following form
\begin{eqnarray}
 T(1+4 \eta +4 \eta ^2)=8 \Delta \eta (1-\eta)^4  \nonumber \\
 T(8+20 \eta +8 \eta ^2)=8 \Delta (1-\eta)^5,
\end{eqnarray}
we obtain the following equation for the critical packing
fraction:
\begin{equation}
  1-5 \eta -20 \eta^2 -12 \eta^3 = 0.
\end{equation}
From this equation one can get the critical packing fraction
$\eta_c=0.12867...$. It is important to emphasize that in
approximation (\ref{3}) the critical density is fully determined
by the hard core diameter $d$ and does not depend on the well
width $\sigma$.

For the critical temperature one obtains:
\begin{equation}
 T_c=\frac{8 \Delta \eta_c (1-\eta_c)^4}{(1+2 \eta_c)^2}.
 \label{crittem}
\end{equation}

Using the value for the critical packing fraction obtained above
one can write $T_c=0.375312 \Delta$. For the system with $\sigma =
1.35d$, which we study here, $T_c=0.548$.

The liquid - gas (LG) transition line can be obtained by the
Maxwell construction. Figs.~\ref{fig:fig5} ((a) and (b)) show the
LG curve in the $\eta-T$ and $P-T$ planes.

\begin{figure}
\includegraphics[width=6cm, height=5cm]{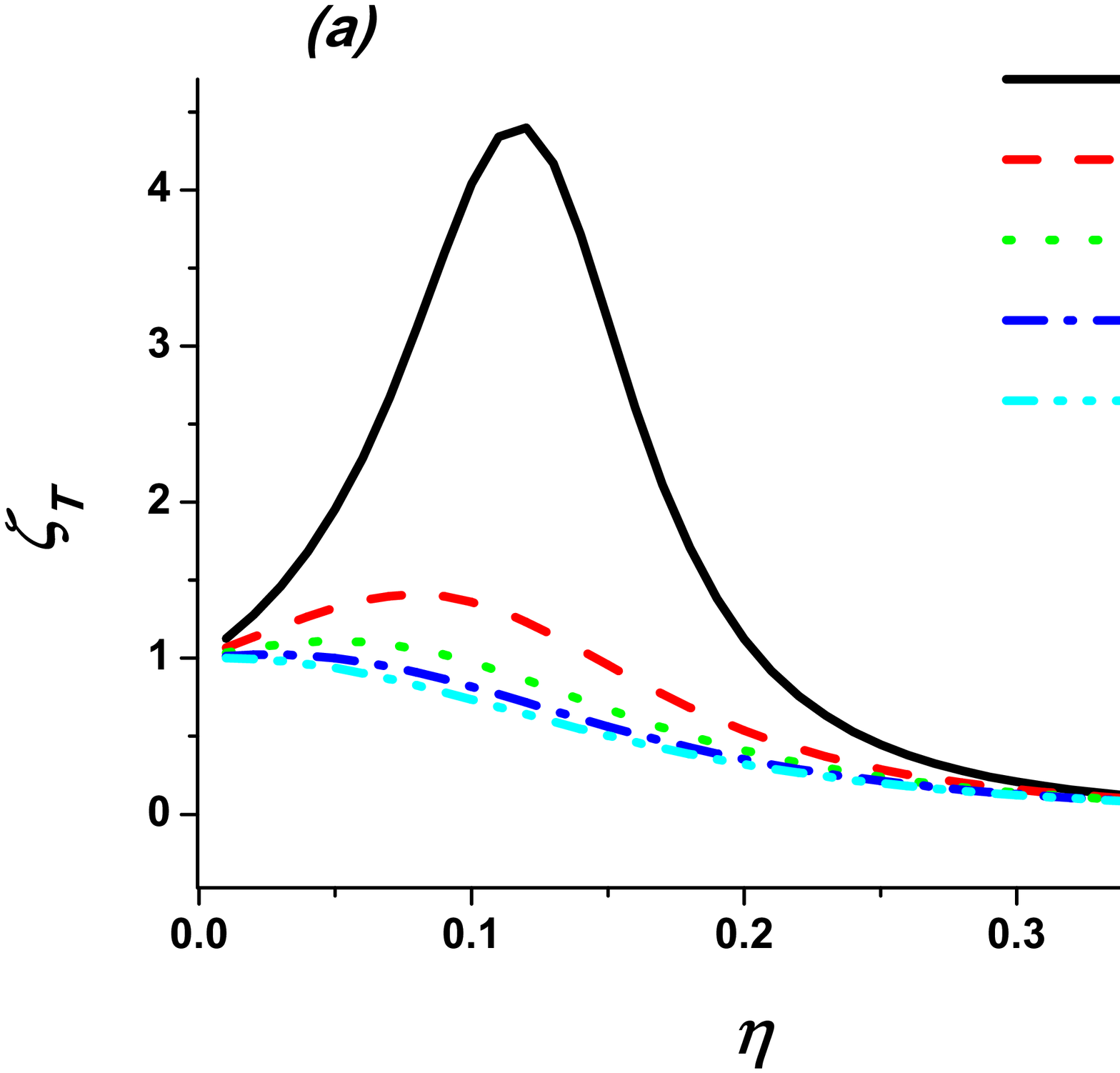}%

\includegraphics[width=6cm, height=5cm]{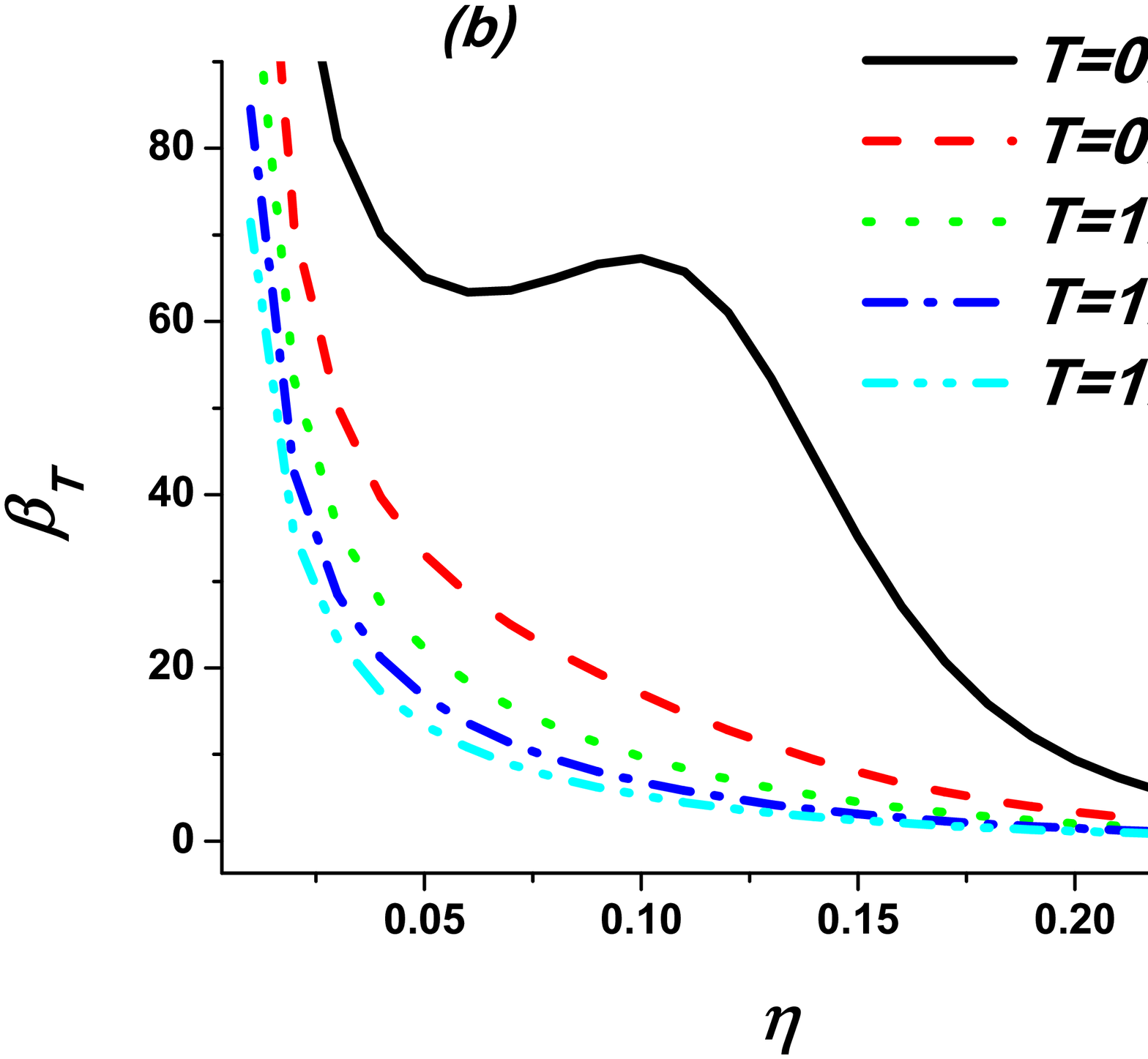}%

\includegraphics[width=6cm, height=5cm]{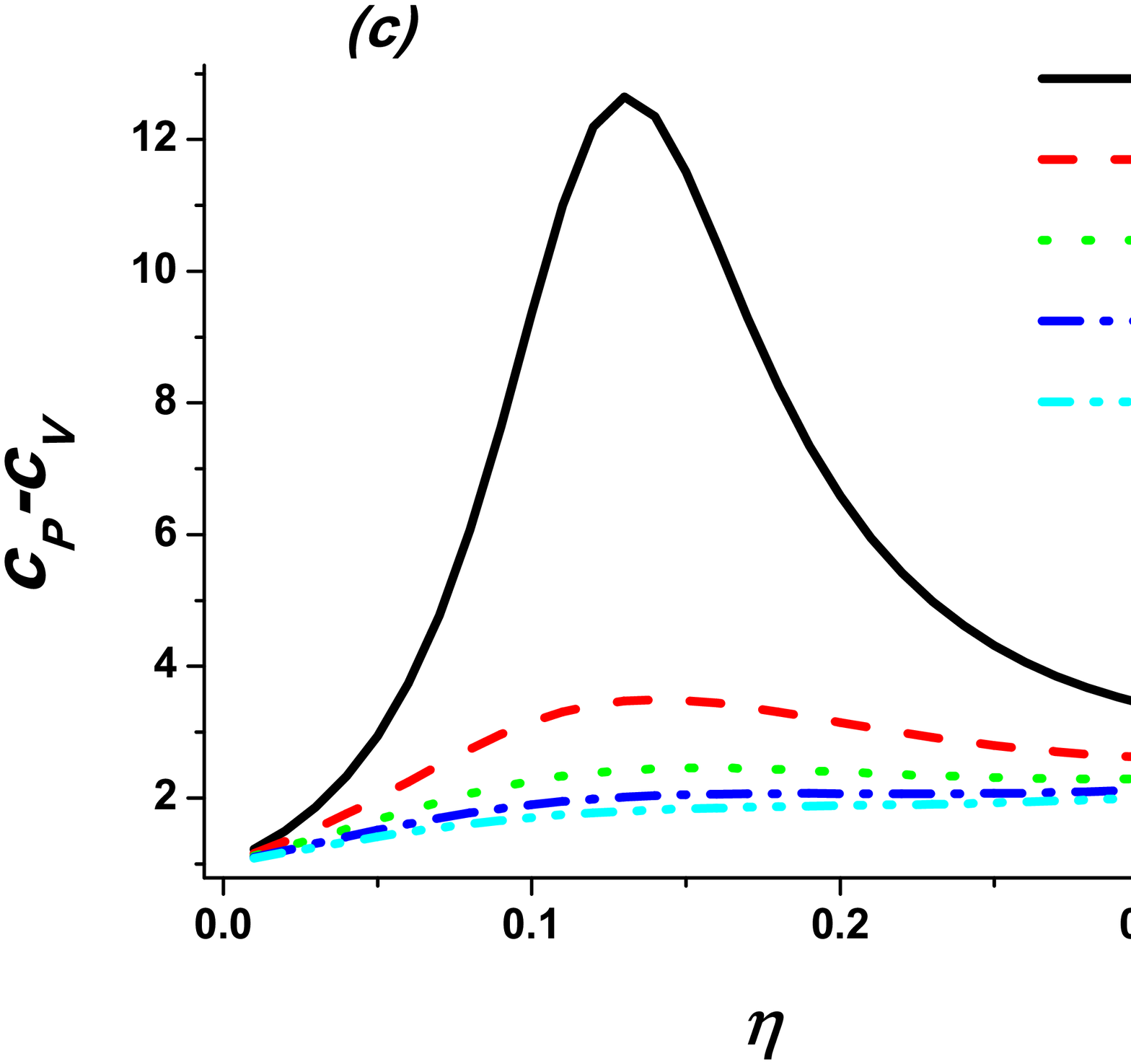}%

\includegraphics[width=6cm, height=5cm]{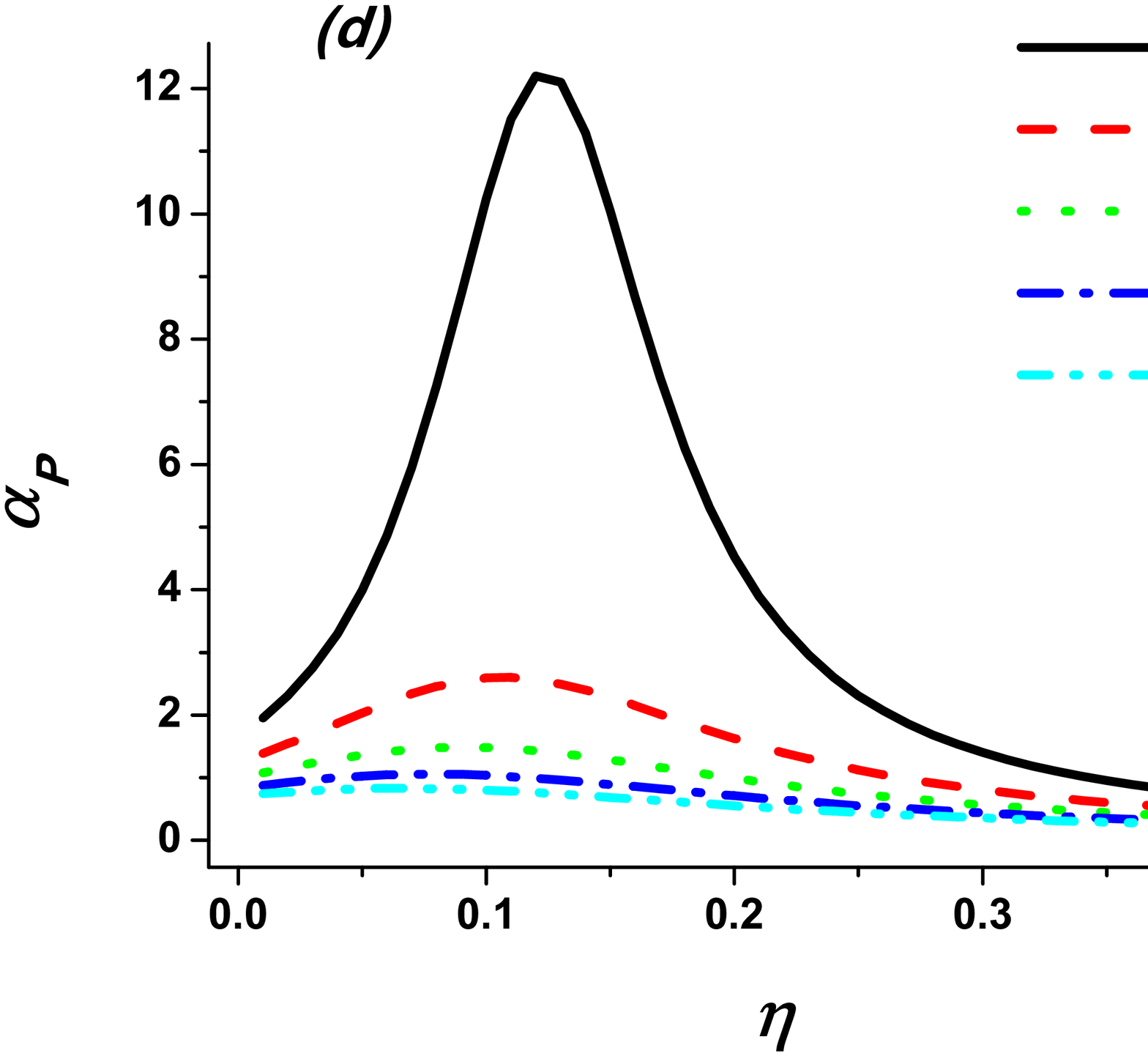}%

\caption{\label{fig:fig3} (Color online) (a) Density fluctuations
$\zeta_T$, (b) isothermal compressibility $\beta_T$, (c) heat
capacity $c_P$, and (d) isobaric thermal expansion coefficient
$\alpha_P$ along several isotherms.}
\end{figure}

It is well known that close to the critical point, many
thermodynamic functions have maxima. Here we calculate the
locations of maxima of different thermodynamic functions in the
framework of the method employed. The real advantage of this
method is that it allows us to calculate the correlation length
$\xi$, i.e. we are able to compare all the definitions of the
Widom line in framework of the purely analytical study of the same
system.

Isothermal density fluctuations are defined as $\zeta _T = T
\left( \frac{\partial \rho}{\partial P} \right ) _T$. From the
equations given above one obtains:
\begin{equation}
  \zeta_T = T \frac{\partial \eta}{\partial P}=\frac{T(1-
  \eta)^4}{T(1+2 \eta)^2-8 \Delta \eta (1-\eta)^4}.
\end{equation}

Fig.~\ref{fig:fig3} (a) shows the density fluctuations along
several isotherms. The $\zeta_T$ maxima along isotherms are
determined from the following equation:
\begin{equation}
  T=\frac{2 \Delta (1- \eta)^5}{2+5 \eta + 2 \eta^2}.
\end{equation}

The isothermal compressibility is defined as $\beta_T
=-\frac{1}{V} \left ( \frac{\partial V} {\partial P} \right ) _T$.
Rewriting it in terms of $\eta$, one obtains:
\begin{equation}
  \beta_T=\frac{1}{\eta} \left ( \frac{\partial \eta}{\partial P}
  \right ) _T= \frac{1}{\eta} \frac{(1- \eta )^4}{T(1+2 \eta)^2 -8 \Delta \eta (1-
  \eta)^4}.
\end{equation}

Fig.~\ref{fig:fig3} (b) shows the compressibilities $\beta_T$
along several isotherms. The corresponding maxima are obtained
from the equation:
\begin{equation}
  T=\frac{16 \Delta \eta (1- \eta)^5}{1+ 11 \eta +20 \eta^2 +4
  \eta^3}.
\end{equation}

The heat capacity $C_P$ can be calculated from the formula:
\begin{equation}
  C_P -C_V=-T \frac{\left ( \frac{\partial P}{\partial T} \right
  )^2_V}{\left ( \frac{\partial P} {\partial V} \right )_T}.
\end{equation}

In terms of $\eta$ this formula is
\begin{equation}
  C_P -C_V=\frac{T(1+\eta+\eta^2)^2}{(1-\eta)^2(T(1+2 \eta)^2-8 \Delta \eta
  (1- \eta)^4)}.
\end{equation}

\begin{figure}
\includegraphics[width=8cm]{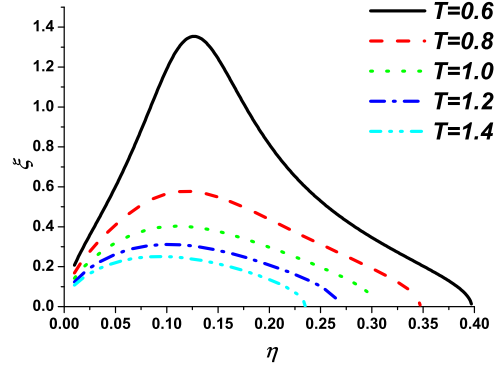}%

\caption{\label{fig:fig4} (Color online) Correlation length $\xi$
along several isotherms.}
\end{figure}

\begin{figure}
\includegraphics[width=8cm]{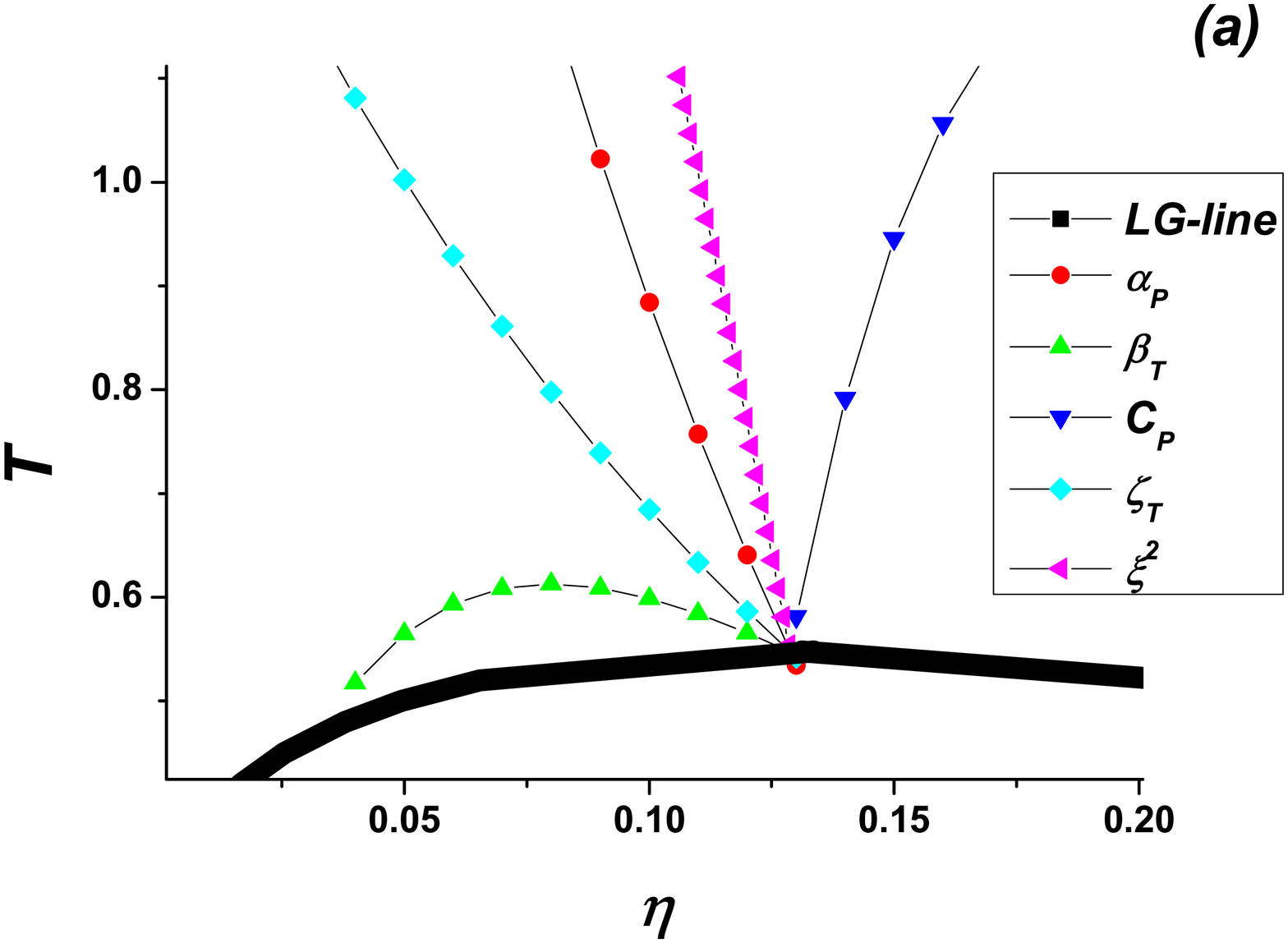}%

\includegraphics[width=8cm]{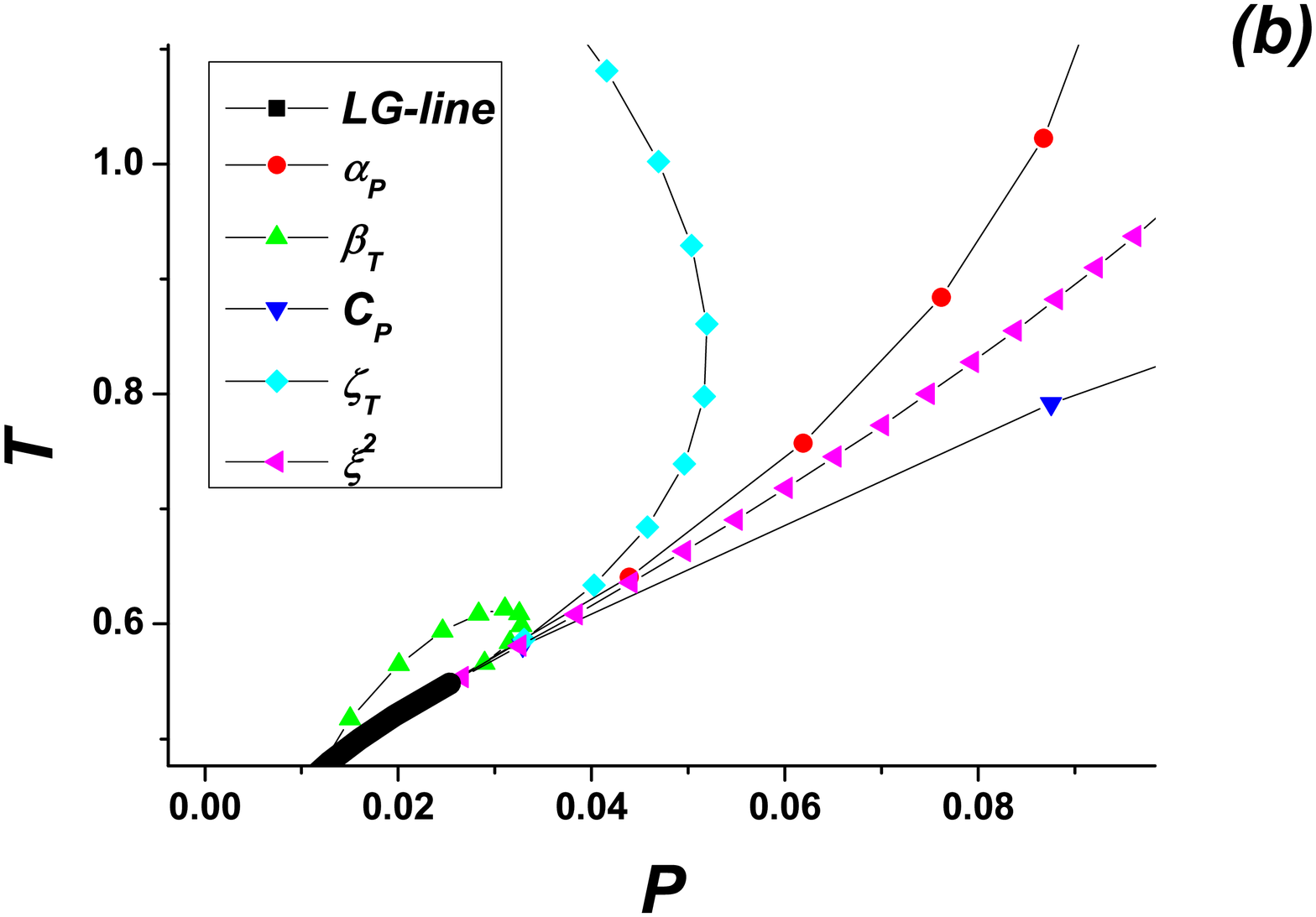}%

\caption{\label{fig:fig5} (Color online) Location of maxima of
different thermodynamic quantities close to the LG curve (a) in
the $\eta-T$ and (b) $P-T$ planes.}
\end{figure}

Some examples of the heat capacities along isotherms are shown in
Fig.~\ref{fig:fig3} (c). The corresponding maxima can be
calculated from the equation:
\begin{equation}
  T=\frac{4 \Delta (1- \eta)^4(-1+8 \eta +8 \eta^2 +3 \eta^3)}{3 \eta (2+5 \eta +2
  \eta^2)}.\label{tcm}
\end{equation}
Taking into account that, as in the case of the vdW model
\cite{stanley_book}, $C_V$ above the critical point  is equal to
the ideal gas value, Eq.~(\ref{tcm}) corresponds to the line of
the supercritical maxima of $C_P$. However, in contrast to the vdW
model, where the line of $C_P$ maxima is located along the
critical isochore \cite{br_jcp}, in this model the supercritical
behavior of $C_P$ (see Fig.~\ref{fig:fig4}) is similar to that in
the case of the LJ fluid \cite{jpcb,may}.

The isobaric thermal expansion coefficient is
$\alpha_P=-\frac{1}{V} \left ( \frac{\partial V}{\partial T}
\right ) _P$. Using the EoS (\ref{eos}), one obtains
\begin{equation}
  \alpha_P= \frac{1}{\eta} \left ( \frac{\partial \eta}{\partial T} \right
  )  _P = \frac{(1+ \eta + \eta^2)(1- \eta)}{T(1+2 \eta)^2 - 8 \Delta \eta (1-
  \eta)^4}.
\end{equation}

Some examples of the $\alpha_P$ behavior along isotherms are shown
in Fig.~\ref{fig:fig3} (d). The corresponding isothermal maxima
are given by the equation
\begin{equation}
  T=\frac{-8 \Delta (1- \eta)^4 (-1 + 4 \eta +4 \eta^2+ 2 \eta^3
  )}{4+ 8 \eta + 3 \eta^2 + 8 \eta^3 + 4 \eta^4}.
\end{equation}

The correlation length can be calculated in the following way
\cite{book}:
\begin{equation}
  \xi^2=\frac{R^2}{1-\rho \tilde{C}_0(T)},
\end{equation}
where
\begin{equation}
 R^2=\frac{\rho}{6} \int c(r)r^2 d\textbf{r},\label{r2}
\end{equation}
and
\begin{equation}
\tilde{C}_0(T)=\int c(r) d\textbf{r}.\label{c0}
\end{equation}

Using the formulas for the direct correlation function
\cite{book}, one can write:
\begin{equation}
  1- \rho \tilde{C}_0(T)=\lambda_1 - \frac{8 \eta \Delta}{T}.
\end{equation}

Substituting this equation into the one for $R^2$, one obtains:
\begin{equation}
 R^2=\frac{\eta d^2}{20} \left [ \frac{-16+11 \eta -4
 \eta^2}{(1-\eta)^4}+ \frac{16\Delta'}{T} \right ],
\end{equation}
where $\Delta'=(\sigma^5-d^5)/d^5$.

Finally, for the correlation length one obtains:
\begin{equation}
  \tilde{\xi}^2=\frac{\xi^2}{d^2}=\frac{1}{20} \frac{T(-16 \eta +11 \eta^2 -4 \eta^3)+
  16\eta\Delta'(1-\eta)^4}{T(1+2 \eta)^2-8 \Delta \eta (1- \eta)^4}.
\end{equation}

Examples of the correlation length along several isotherms are
given in Fig.~\ref{fig:fig4}. The maxima of the correlation length
can be calculated as the solutions of the following equation:
\begin{eqnarray}
 & & 4\Delta \eta^2 (1-\eta)^3 (-53+25\eta-8\eta^2)+ \nonumber\\
 &+& 8 \Delta'(1-\eta)^3
(-1+5\eta+20\eta^2+12\eta^3)+ \nonumber\\
 &+&\ T(8-11 \eta - 48 \eta^2 + 16
\eta^3 + 8\eta^4) =0. \label{27}
\end{eqnarray}


Fig.~\ref{fig:fig5}(a) shows the LG curve and the points of maxima
of all the quantities described above. One can see, that the
curves of maxima of different thermodynamic functions quickly
diverge, and even rather close to the critical point, one cannot
consider the location of the maxima as a single curve. They
actually represent a bunch of curves in the $\eta-T$ plane.

In particular, one can see that the correlation length maxima are
located between the $\alpha_P$ and $C_P$ maxima, and even the
qualitative behavior of the maxima of correlation length is
opposite to that of the heat capacity in the $\eta-T$ plane.

In Fig.~\ref{fig:fig5}(b), the behaviors of maxima of the
isothermal compressibility $\beta_T$, isobaric heat capacity
$C_P$, density fluctuations $\zeta_T$, thermal expansion
coefficient $\alpha_T$, and correlation length $\xi$ at constant
temperature are shown in the $P-T$ plane. One can see again, that
the bunch of ridges merges into a single line in the very vicinity
of the critical point and widens rapidly upon departure from the
critical point. If the Widom line is defined with the help of the
correlation function $\xi$, then the Widom line  will not follow
the slope of any response function extrema, except in the very
close vicinity of the critical point.

It is interesting to note that the sequence of the $\xi$,
$\beta_T$, $C_P$,  $\zeta_T$, and $\alpha_T$ extrema in
Fig.~\ref{fig:fig5} is the same as the corresponding sequence for
the vdW and LJ fluids \cite{jpcb,br_jcp,may}. It seems that the
similar behavior of the correlation length maxima line in our
model and the corresponding lines obtained in \cite{may} may be
considered as an evidence of the universality of a particular
location of the "true" Widom line.

It seems that the Widom line defined in this way cannot be used as
a single boundary that separates the supercritical region into the
gaslike and liquidlike regions. For this purpose, a new dynamic
line on the phase diagram in the supercritical region, namely the
Frenkel line has been proposed recently
\cite{br_ufn,br_jl,br_pre,br_jcp1,br_prl}. The intersection of
this line corresponds to radical changes of system properties.
Liquids in this region exist in two qualitatively different
states: "rigid" and "nonrigid" liquids. The rigid to nonrigid
transition corresponds to the condition $\tau\approx \tau_0$,
where $\tau$ is the liquid relaxation time and $\tau_0$ is the
minimal period of transverse quasiharmonic waves. This condition
defines a new dynamic crossover line on the phase diagram and
corresponds to the loss of shear stiffness of a liquid at all
available frequencies and, consequently, to qualitative changes in
many properties of the liquid. In contrast to the Widom line that
exists only near the critical point, the new dynamic line is
universal. It separates two liquid states at arbitrarily high
pressure and temperature and exists in systems where the
liquid-gas transition and the critical point are absent
altogether. The location of the line can be rigorously and
quantitatively established on the basis of the velocity
autocorrelation function and mean-square displacements. It was
also shown that the positive sound dispersion disappears in the
vicinity of the Frenkel line
\cite{bryk2,br_jl,br_pre,br_jcp1,br_prl}.

In conclusion, in the present paper we propose the van der
Waals-like model which allows a purely analytical study of fluid
properties including the equation of state, phase behavior and
supercritical fluctuations. We take a square-well system as an
example and calculate its liquid - gas transition line and
supercritical fluctuations. Employing this model allows us to
calculate the correlation length in the fluid $\xi$, isothermal
compressibility $\beta_T$, the isobaric heat capacity $C_P$,
density fluctuations $\zeta_T$, and thermal expansion coefficient
$\alpha_T$. It is shown that, in accordance with our recent
results obtained for Lennard-Jones and van der Waals liquids
\cite{jpcb,br_jcp}, the bunch of extrema merges into a single line
in the very close vicinity of the critical point and widens
rapidly upon departure from the critical point. If the "true"
Widom line is defined with the aid of the correlation function
$\xi$, one can see that the Widom line  does not follow the slope
of any response function extrema except those located in the very
close vicinity of the critical point. It seems that the Widom line
defined in this way cannot be used as the boundary that separates
the supercritical region into the gaslike and liquidlike regions.
As it has been shown recently, a new dynamic line on the phase
diagram in the supercritical region, namely the Frenkel line, can
be used for this purpose
\cite{br_ufn,br_jl,br_pre,br_jcp1,br_prl}.

\bigskip

\begin{acknowledgments}
We are greateful to S. M. Stishov and A.G. Lyapin for stimulating
discussions. The work was supported in part by the Russian
Foundation for Basic Research (Grants No 14-02-00451, 13-02-12008,
13-02-00579, and 13-02-00913) and the Ministry of Education and
Science of Russian Federation (project MK-2099.2013.2).
\end{acknowledgments}


\end{document}